\documentclass[twoside]{article}
\usepackage{fleqn,espcrc2}
\usepackage{epsf}

\newcommand{\AmS}{{\protect\the\textfont2
  A\kern-.1667em\lower.5ex\hbox{M}\kern-.125emS}}

\font\twelvebb=msbm10 scaled 1200
\font\tenbb=msbm10
  \newfam\bbfam
  \def\bb{\fam\bbfam\twelvebb}
  \textfont\bbfam=\twelvebb
  \scriptfont\bbfam=\tenbb
  \scriptscriptfont\bbfam=\scriptfont\bbfam
\font\twelveeusm=eusm10 scaled 1200
\font\teneusm=eusm10
  \newfam\eusmfam
  
  \textfont\eusmfam=\twelveeusm
  \scriptfont\eusmfam=\teneusm
  \scriptscriptfont\eusmfam=\scriptfont\eusmfam
\font\twelvefk=eufm10 scaled 1200
\font\tenfk=eufm10
  \newfam\fkfam
  \def\fk{\fam\fkfam\twelvefk}
  \textfont\fkfam=\twelvefk
  \scriptfont\fkfam=\tenfk
  \scriptscriptfont\fkfam=\scriptfont\fkfam
\def\draftversion{N}		    
\def\note[#1]#2{\message{(#1)}\if\draftversion Y{\noindent\em[#2]\/}\fi}
\if \draftversion Y
\fi
\def\erf{\mathop{\rm erf}}	    
\def\Z{{\bb Z}}			    
\def\opt#1{{{#1}_{\hbox{\rm\tiny opt}}}}  
\def\expo#1{{{#1}_{\hbox{\rm\tiny exp}}}} 
\def\acc#1{{{#1}_{\hbox{\rm\tiny acc}}}}  
\def\n{N}			    
\def\dt{{\delta\tau}}		    
\def\pmomtheta{\vartheta}	    
\def\pmomthetaopt{\opt\pmomtheta}   
\def\trjlen{\tau}                   
\def\trjlenav{{\bar\trjlen}}	    
\def\nexp{\expo\n}		    
\def\dH{{\delta H}}		    
\def\pacc{\acc P}		    
\def\paccav{\acc{\bar P}}	    
\def\cost{{\fk C}}		    
\def\hmc{Hybrid Monte Carlo}	    
\def\twovector[#1,#2]{\left(\begin{array}{c} #1 \\ #2 \end{array}\right)}
\def\twomatrix[#1,#2;#3,#4]%
  {\left(\begin{array}{cc}
    #1 & #2 \\
    #3 & #4
  \end{array}\right)}
\def\threevector[#1,#2,#3]%
  {\left(\begin{array}{c} #1 \\ #2 \\ #3\end{array}\right)}
\def\sixvector[#1,#2,#3,#4,#5,#6]%
  {\left(\begin{array}{c} #1 \\ #2 \\ #3\\ #4\\ #5\\ #6\end{array}\right)}
\def\threematrix[#1,#2,#3;#4,#5,#6;#7,#8,#9]%
  {\left(\begin{array}{ccc}
    #1 & #2 & #3 \\
    #4 & #5 & #6 \\
    #7 & #8 & #9
  \end{array}\right)}
\def\cuberoot#1{\root3\of{#1}}	    
\def\fifthroot#1{\root5\of{#1}}	    
\def\stripsign#1{\if+#1\else#1\fi}  
\def\crte(#1,#2,#3){\left(\stripsign#3#2\cuberoot2#1\cuberoot4\right)}
\def\frte(#1,#2,#3,#4,#5){{\left(
  \begin{array}{c}
    \scriptstyle{\stripsign#5} \\
    \scriptstyle{#4\fifthroot2} \\
    \scriptstyle{#3\fifthroot4} \\
    \scriptstyle{#2\fifthroot8} \\
    \scriptstyle{#1\fifthroot{16}}
  \end{array}
\right)}}
\def\notbeta#1{\if\beta#1\else(#1)\fi}
\def\G(#1){G\notbeta#1}
\def\GC_#1(#2){{G^{(C)}_{#1}\notbeta#2}}
\def\GS_#1(#2){{G^{(S)}_{#1}\notbeta#2}}
\def\GCC_#1(#2){{G^{(CC)}_{#1}\notbeta#2}}
\title{Cost of Generalised HMC Algorithms for Free Field Theory}

\author{A.~D.~Kennedy and Brian Pendleton
	\address{Department of Physics and Astronomy, 
	The University of Edinburgh,
	Edinburgh, EH9~3JZ, Scotland}}

\begin{document}

\begin{abstract}
	We study analytically the computational cost of the Generalised
Hybrid Monte Carlo (GHMC) algorithm for free field theory. We
calculate the autocorrelation functions of operators quadratic in the
fields, and optimise the GHMC momentum mixing angle, the trajectory
length, and the integration stepsize. We show that long trajectories
are optimal for GHMC, and that standard HMC is much more efficient
than algorithms based on the Second Order Langevin (L2MC) or Kramers
Equation. We show that contrary to naive expectations HMC and L2MC
have the same volume dependence, but their dynamical critical exponents
are $z = 1$ and $z = 3/2$ respectively.
\vspace{1pc}
\end{abstract}

\maketitle

\section{GENERALISED HMC}

The work reported here extends results presented
previously~\cite{kennedy90,kennedy95a}, to which the reader is
referred for details.  We begin by recalling that the generalised
HMC~\cite{kennedy95a} algorithm is constructed from two kinds of
update for a set of fields $\phi$ and their conjugate momenta $\pi$.

\textbf{Molecular Dynamics Monte Carlo:}
	This consists of three parts: (1)~\emph{MD:} an approximate
integration of Hamilton's equations on phase space which is exactly area-preserving
and reversible; $U(\trjlen): (\phi,\pi) \mapsto (\phi',\pi')$ where
$\det U=1$ and $U(\trjlen) = U(-\trjlen)^{-1}$.  (2)~A momentum flip
$F: \pi\mapsto-\pi$.  (3)~\emph{MC:} a Metropolis accept/reject test.

\textbf{Partial Momentum Refreshment:}
	This mixes the Gaussian-distributed momenta $\pi$ with
Gaussian noise $\xi$:
\begin{displaymath}
  \twovector[\pi',\xi'] =
    \twomatrix[ \cos\pmomtheta, \sin\pmomtheta;
			 -\sin\pmomtheta, \cos\pmomtheta]
      \cdot F \twovector[\pi,\xi]
\end{displaymath} The HMC algorithm~\cite{duane87} is the special case
where $\pmomtheta= {\pi\over2}$.  The L2MC/Kramers algorithm
\cite{horowitz90a,beccaria94a} corresponds to choosing arbitrary
$\pmomtheta$ but MDMC trajectories of a single leapfrog integration
step.

\subsection{Tunable Parameters}
\label{sec:tunable-parameters}

The GHMC algorithm has three free parameters, the trajectory length
$\trjlen$, the momentum mixing angle $\pmomtheta$, and the integration
step size~$\dt$.  These may be chosen arbitrarily without affecting
the validity of the method, except for some special values for which
the algorithm ceases to be ergodic. We may adjust these parameters to
minimise the cost of a Monte Carlo computation, and the main goal of
this work is to carry out this optimisation procedure for free field
theory.

Horowitz pointed out that the L2MC algorithm has the advantage of
having a higher acceptance rate than HMC for a given step size, but he
did not take in to account that it also requires a higher acceptance
rate to get the same autocorrelations because the trajectory is
reversed at each MC rejection. It is not obvious a priori which of
these effects dominates.

The parameters $\trjlen$ and $\pmomtheta$ may be chosen independently
from some distributions $P_R(\trjlen)$ and $P_M(\pmomtheta)$ for each
trajectory. In the following we shall consider various choices for the
momentum refreshment distribution $P_M$, but we shall always take a
fixed value for $\pmomtheta$.

\section{FREE FIELD THEORY}
Consider a system of harmonic oscillators $\{\phi_p\}$ for $p\in\Z_V$.
The Hamiltonian on phase space is $H = {1\over2}\sum_{p\in\Z_V}
\left(\pi_p^2 + \omega_p^2\phi^2\right)$. This describes free field
theory in momentum space if the frequencies $\omega_p$ are chosen as
\begin{equation}
  \omega_p^2 = m^2 + 4 \sum_{\mu=1}^d \sin^2 \frac{\pi p_\mu}{L}.
  \label{eq:frequency-definition}
\end{equation}

\section{AUTOCORRELATION FUNCTIONS}

\subsection{Simple Markov Processes}
\label{simple-markov-processes}

Let $(\phi_1,\phi_2,\ldots,\phi_N)$ be a sequence of field configurations
generated by an equilibrated ergodic Markov process, and let $\langle
\Omega(\phi) \rangle$ denote the expectation value of some connected operator
$\Omega$. We may define an \emph{unbiased estimator} $\bar\Omega$ over the
finite sequence of configurations by $\bar\Omega \equiv {1\over N} \sum_{t=1}^N
\Omega(\phi_t)$, As usual, we define $C_\Omega(\ell) \equiv {\bigl\langle
\Omega(\phi_t+\ell)\Omega(\phi_{t}) \bigr\rangle \over \bigl\langle
\Omega(\phi)^2 \bigr\rangle}$ as the \emph{autocorrelation function}
for~$\Omega$. The variance of the estimator $\bar\Omega$ is
\begin{displaymath}
  \langle{\bar\Omega}^2\rangle = \{1 + 2A_\Omega\}
    {\bigl\langle \Omega(\phi)^2 \bigr\rangle \over N}
    \left[1 + O\left({\nexp\over N}\right)\right],
\end{displaymath}
where $A_\Omega\equiv\sum_{\ell=1}^\infty C_\Omega(\ell)$ is the
\emph{integrated autocorrelation function} for the operator~$\Omega$ and $\nexp$
is the \emph{exponential autocorrelation time}. This result tells us that on
average $1+2A_\Omega$ correlated measurements are needed to reduce the variance
by the same amount as a single truly independent measurement.

\subsection{Autocorrelation Functions for Quadratic Operators}

In order to carry out these calculations we make the simplifying
assumption that the acceptance probability
$\min\bigl(1,e^{-\dH}\bigr)$ for each
trajectory may be replaced by its value averaged over phase space
$\pacc\equiv \left\langle \min\bigl(1, e^{-\dH}\bigr)
\right\rangle$; we neglect correlations between successive
trajectories. Including such correlations leads to seemingly
intractable complications. It is not obvious that our assumption
corresponds to any systematic approximation except, of course, that it
is valid when $\dH=0$.
	
	Details of the calculation of $\pacc$ for leapfrog integration
are published elsewhere~\cite{kennedy90,kennedy95a,kennedy99}.

\section{COMPARISON OF COSTS}

If we make the reasonable assumption that the cost of the computation
is proportional to the total fictitious (MD) time for which we have to
integrate Hamilton's equations, then the cost $\cost$ per independent
configuration is proportional to $(1+2A_\Omega) \trjlenav/\dt$ with
$\trjlenav$ denoting the average length of a trajectory.  The optimal
trajectory length is obtained by minimising the cost, that is by
choosing $\trjlenav$ so as to satisfy ${d\cost/d\trjlenav} =
{d\cost/d\pmomtheta} = 0$.

	We wish to compare the performance of the HMC, L2MC and GHMC
algorithms for one dimensional free field theory. To do this we
compare the cost of generating a statistically independent measurement
of the magnetic susceptibility $M^2_c$, choosing the optimal values
for the angle $\pmomtheta$ and the average trajectory length
$\trjlenav$.  We can minimise the cost with respect to $\pmomtheta$
without having to specify the form of the refresh distribution.

The next step is to minimise the cost with respect to the average
trajectory length $\xi=\omega\trjlenav$. Strictly speaking we should
note that the acceptance probability $\paccav$ is a function of
$\trjlenav$, but to a good approximation we may assume that $\pacc$
depends only upon the integration step size $\dt$ \emph{except} in the
case of very short trajectories.

\subsection{Exponentially Distributed Trajectory Lengths}

	To proceed further we need to choose a specific form for the
momentum refresh distribution. In this section we will present results
for the case of exponentially distributed trajectory lengths,
$P_R(\trjlen) = re^{-r\trjlen}$ where the parameter $r$ is just the
inverse average trajectory length $r=1/\trjlenav$.

\begin{figure}[t]
  \begin{center}
    \epsfxsize=0.5\textwidth 
	\leavevmode\epsffile{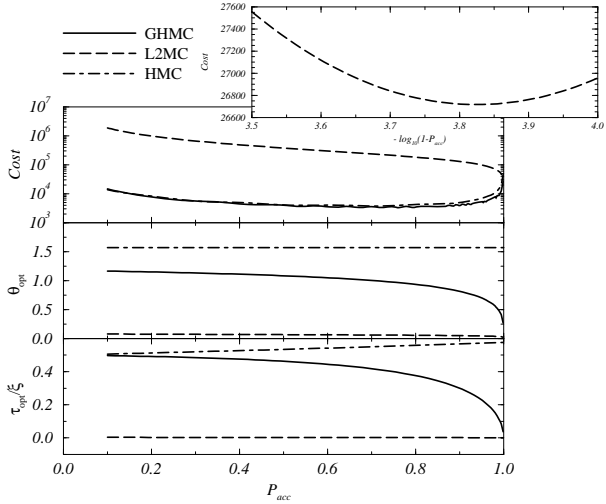}
  \end{center}
  \def\1{{\mathop{\rm opt}}}
  \caption[laplace-opt.eps]{Cost as a function of average Metropolis
    acceptance rate for the GHMC algorithm compared to HMC and L2MC for free
    field theory. The operator under consideration is the ``magnetic
    susceptibility'', i.e., the connected quadratic operator depending only on
    the lowest frequency mode. The corresponding parameters, the momentum
    mixing angle $\pmomtheta_\1$ and the average trajectory length measured as
    a fraction of the correlation length $\trjlen_\1/\xi$ are also shown,
    all as a function of the acceptance rate $\paccav$. The inset graph shows
    the region where the acceptance rate is very close to unity which is where
    the L2MC algorithm has its minimum cost.}
  \label{fig:laplace-opt}
\end{figure}

The cost at the point $(\opt{c}\equiv\cos\opt{\pmomtheta},\opt{\xi})$
is
\[
\cost^{\hbox{\tiny exp}}_{M^2_c} (\dt, \pmomthetaopt, \opt{\xi}) ~ = ~
\]
\vspace*{-3ex}
\[
  \frac 
   {
   \left(
   \begin{array}{c}
	(7\paccav-3\paccav^2-4)\opt{\xi}^2\opt{c}^3 
     + 1 - \opt{c}^2 \\
     {}
     + (-2\paccav+1)\opt{c}
     + (2\paccav-1)\opt{c}^3 \\
     {}
     + (\paccav^2-5\paccav+4)\opt{c}\opt{\xi}^2 \\
     {}
     + (-\paccav+4)\opt{\xi}^2
     + (-4+3\paccav)\opt{c}^2\opt{\xi}^2
   \end{array}
   \right)}
   {
   \left(
   \begin{array}{c}
   \paccav\dt\omega(\paccav-1)\opt{c}^3\opt{\xi} \\
     {}
     + \opt{\xi}\paccav\dt\omega 
     - \opt{\xi}\paccav\dt\omega{\opt{c}}^2 \\
     {}
     - \paccav\dt\omega(-1+\paccav)\opt{c}\opt{\xi}
   \end{array}
   \right)} \, .
\]
	This solution is a function of $\dt$ and $\paccav$ which are
not independent variables, and using the results for
$\paccav(\trjlen,\dt)$~\cite{kennedy90,kennedy95a} we can compute the
cost as a function of $\paccav$ as shown in
Figure~\ref{fig:laplace-opt}.

\subsubsection{HMC}

The {\hmc} algorithm corresponds to setting $\pmomtheta=\pi/2$, and we
find that the optimal trajectory length in this case is
$\opt\xi=1/\sqrt{4-\paccav}$, corresponding to a cost
\[
  \opt\cost = {2\sqrt{4-\paccav}\over\paccav\dt\omega}.
\]
This is also shown in Figure~\ref{fig:laplace-opt}.

\subsection{Fixed Length Trajectories}

For fixed length trajectories we shall only analyse the case of L2MC
for which the trajectory length $\xi=\omega\dt$. In this case the
value of $\opt{\pmomtheta}$ and the corresponding cost are also
plotted in Figure~\ref{fig:laplace-opt}. From this figure it is clear
that the minimum cost occurs for $\paccav$ very close to unity, where
the scaling variable $x=V\dt^6$ is very small. We may then express
$\opt{c}$ and $\paccav$ as power series in $x$, keeping only the first
few terms.  From these relations we find that the minimum cost for
L2MC is
\[
  \opt{\cost} = \left(10\over\pi\right)^{1/4} V^{5/4} m^{-3/2}.
\]
	This result tells us that not only does the tuned L2MC
algorithm have a dynamical critical exponent $z=3/2$, but also it has
a volume dependence of exactly the same form as
HMC~\cite{kennedy90,kennedy95a}. We may understand why this behaviour
occurs rather than the naive $V^{7/6}m^{-1}$ by the following simple
argument.

	If $\paccav<1$ then the system will carry out a random walk
backwards and forwards along a trajectory because the momentum, and
thus the direction of travel, must be reversed upon a Metropolis
rejection. A simple minded analysis is that the average time between
rejections must be $O(1/m)$ in order to achieve $z=1$. This time is
approximately
\[
  \sum_{n=0}^\infty \paccav^n (1-\paccav) n\dt = {\paccav\dt\over1-\paccav}
  = {1\over m}.
\]
	For small $\dt$ we have $1-\paccav =
\erf\sqrt{kV\dt^6}\propto\sqrt{V\dt^6}$ where $k$ is a constant, and
hence we must scale $\dt$ so as to keep $V\dt^4/m^2$ fixed. Since the
L2MC algorithm has a naive dynamical critical exponent $z=1$,
this means that the cost should vary as $\cost\propto
V(V\dt^4m^{-2})^{1/4} / m\dt = V^{5/4} m^{-3/2}$.

\bibliographystyle{unsrt}

\end{document}